\documentclass[12pt]{article}
\begin{document}

\addtolength{\baselineskip}{0.5\baselineskip}

\title{\textbf{Single-Particle $\it{Green}$ Function Approach and
Correlated Atomic or Molecular Orbitals}}
\author{Liqiang Wei\\
Institute for Theoretical Atomic, Molecular and Optical Physics\\
Harvard University, Cambridge, MA 02138\\
$\it{Email: liqiangjohnwei@yahoo.com}$}

\maketitle

\begin{abstract}
\vspace{0.05in} In this paper, we propose a generic and systematic
approach for study of the electronic structure for atoms or
molecules. In particular, we address the issue of single particle
states, or orbitals, which should be one of the most important
aspects of a quantum many-body theory. We argue that the
single-particle $\it{Green}$ function provides a most general
scheme for generating these single particle states or orbitals. We
call them the $\it{correlated}$ atomic or molecular orbitals to
make a distinction from those determined from $\it{Hartree-Fock}$
equation. We present the calculation of the single particle
properties (i.e., the electron affinities $(EA's)$ and ionization
potentials $(IP's)$) for the $H_{2}O$ molecule using the
correlated molecular orbitals in the context of quantum chemistry
with a second-order self energy. We also calculate the total
ground state energy with a single $Slater$ wavefunction determined
only from the hole states. Comparisons are made with available
experimental data as well as with those from the
$\it{Hartree-Fock}$ or density functional theory $(DFT)$
calculations. We conclude that the correlated atomic or molecular
orbital approach provides a strictest and most powerful method for
studying the single-particle properties of atoms or molecules. It
also gives a better total energy than do the $\it{Hartree-Fock}$
and $\it{DFT}$ even at the single $\it{Slater}$ determinant level.
It promises that a correlation theory based on the correlated
atomic or molecular orbitals will become an approach which
possesses the advantages and also overcomes their shortcomings of
current quantum chemistry methods based on either the conventional
quantum many-body theory or the $DFT$.

\end{abstract}

\vspace{0.35in}
\section{Introduction}

 The single particle approximation, or the concept of atomic or molecular
orbitals in the context of quantum chemistry, is a natural and
almost a necessary scenario for solving an interacting
many-electron system for atoms, molecules, or
solids~\cite{mulliken}. This is a reflection of not only a
physical existence but also possibly a mathematical reality. The
usual equation which is being used to determine the orbitals is
the $\it{Hartree-Fock}$ equation~\cite{hartree,fock}. The rest of
endeavor to remedy the approximation resulting from a replacement
of the whole many-body wavefunction by a single determinant used
in the $\it{HF}$ scheme is called the correlation issue. This is a
most difficult problem and constitutes the major activity of
researches for the quantum chemistry community in the last 50
years~\cite{keinan}. According to the energy scale principle we
described in paper~\cite{wei1}, the $\it{Hartree-Fock}$ scheme
should be a good approximation when the single determinant
wavefunction dominates and there is no any significant mixing with
the nearby configurations. This is typically the case when we
compute the energetics for molecules with a stable geometric
structure. The subsequent perturbation correction for the
correlation such as $\it{MPPT}$ is also proved to be
powerful~\cite{pople1}. However, there are the situations when the
configuration mixing is a prominent or dominant phenomenon, and
the description with more than one single configuration seems
necessary. This includes the calculation of transition states or
excited states, and for open-shell molecules, etc. The computation
based on the $\it{Hartree-Fock}$ equation has shown to be
insufficient, and the corresponding perturbation correction has
proved not to be convergent~\cite{schaefer,stillinger,jorgensen}.
The $\it{MCSCF}$ approaches have been introduced to investigate
this type of nondynamic or static correlation issue, and they have
become one of the most popular approaches for the study of
molecular electron correlation~\cite{wei1,das,gilbert,roos}.
However, the size of molecular systems that this type of
approaches can address are still limited because of the
difficulties in selecting the appropriate configuration states and
in achieving the convergence to the correct state of the
interest~\cite{roos}. Another important and significant advance in
the fields of electron correlation is the development of density
function theory ($\it{DFT}$)~\cite{kohn,parr}. Instead of working
with a multi-configurational framework, it intents to incorporate
the exchange-correlation effect into a single-particle potential
formalism. It has already shown its very usefulness in the study
of the electronic structure for large systems with utilization of
relatively smaller computational efforts. Nevertheless, there
exist some serious drawbacks for the method when seen either from
theoretical consideration or from the practical performance in
calculation. One shortcoming is that the theory can only study the
ground state problem, and cannot treat the same eigenstate problem
for excited states within one theoretical framework. Another
serious problem is that the actual form for the
exchange-correlation is unknown, or the theory itself gives no
clue for how to approach it. Moreover, the approach fails to or
can not do the accurate computation for the points or situations
when the configuration mixing is
important~\cite{jursic,davidson2,koch}. Indeed, it should be a
very difficult thing, intending to replace the intrinsic
$\it{many}$-body effects such as static correlation or
configuration mixing with a $\it{single}$-particle formalism.

Recently, we have demonstrated that a general quantum many-body
perturbation theory can not only be used for understanding the
various electronic phenomena including the nature of chemical
bonds but also serve as a unified theme for constructing general
electronic structure theories and calculation schemes. This also
includes the study of important issues of electron
correlation~\cite{wei1}. This pinpoints the direction and paves
the way for the future investigation. In this paper, we add
another important ingredient to the field of electron correlation
or electronic structure theory in general. We emphasize our
investigation on the issue of single particle approximation, or
the atomic or molecular orbitals for the quantum chemistry
calculation. From the perturbation point of view, this corresponds
to defining a reference Hamiltonian~\cite{wei1,szabo}. We will
show that there exists a strict theoretical formalism, called the
single-particle $\it{Green}$ function, which provides a most
general scheme for generating or determining these single-particle
states up to present time. The theory of single-particle
$\it{Green}$ function has been developed for a long time and used
in many different ways but its full physical meaning or context is
not totally understood or appreciated. This paper aims at a
beginning for a systematic investigation of electron correlation
based on the single-particle $\it{Green}$ function formalism and
within the quantum many-body perturbation theory~\cite{wei1}. In
the next Section, we present its definitions and equations in both
time and energy domains. In particular, we give an energy
eigenequation that solves the single-particle states. We analyze
its intrinsic structure and compare it with other methods. In
Section 3, we calculate both the single-particle properties and
the total energies for the $H_{2}O$ molecule using the
$\it{Hartree-Fock}$, $\it{DFT}$, and correlated molecular orbital
approaches. In the final Section, we analyze and discuss our
results for the calculations and also do the comparison with each
other including the corresponding experimental data. We also
propose a generic electronic structure theory and outline the
future research.

\vspace{0.35in}
\section{Theory}

Two time $(t,t^{'})$ and single-particle (or hole) $\it{Green}$
function is defined as~\cite{schwinger,hedin}
\begin{equation}
G(\vec{x}t,\vec{x}^{'}t^{'})=-i\langle\Psi_{0}|T\{\hat{\psi}(\vec{x},t)
\hat{\psi}^{+}(\vec{x}^{'},t^{'})\}|\Psi_{0}\rangle,
\end{equation}
where $T$ is $\it{Wick}$ time-ordering operator, and
$\hat{\psi}(\vec{x},t)$ and $\hat{\psi}^{+}(\vec{x}^{'},t^{'})$
are the field operators in the $\it{Heisenberg}$ picture
associated with the coordinates $\vec{x}$, which includes both
spatial $\vec{r}$ and spin $\chi$ degrees of freedom. The
$|\Psi_{0}\rangle$ is the exact ground state of an $N$-electron
system being studied. Its Hamiltonian in the field operator
representation can be written as
\begin{equation}
H=\int\hat{\psi}^{+}(\vec{x})h(\vec{x})\hat{\psi}(\vec{x})d\vec{x}+\frac{1}{2}
\int\hat{\psi}^{+}(\vec{x})\hat{\psi}^{+}(\vec{x}^{'})v(\vec{r},\vec{r}^{'})
\hat{\psi}(\vec{x}^{'})\hat{\psi}(\vec{x})d\vec{x}d\vec{x}^{'},
\end{equation}
where the one-body operator $h(\vec{x})$ is the sum of the
electronic kinetic energy operator and its interaction with the
nucleus
\begin{equation}
h(\vec{x}) = -\frac{\hbar^{2}}{2m}\nabla^{2}-\sum_{p}Z_{p}
v(\vec{x}, \vec{R}_{p}),
\end{equation}
and the two-body operator $v(\vec{r},\vec{r}^{'})$ is the
$\it{Coulomb}$ potential
\begin{equation}
v(\vec{r},\vec{r}^{'})=\frac{1}{|\vec{r}-\vec{r}^{'}|}.
\end{equation}
In the energy domain, the $\it{Green}$ function takes the form
\begin{equation}
 G(\vec{x},\vec{x}^{'};\omega)=\sum_{n}\frac{\phi_{n}(\vec{x})\phi^{*}_{n}(\vec{x}^{'})}
 {\omega-\epsilon_{n}},
\end{equation}
where
\begin{equation}
\phi_{n}(\vec{x})=\langle\Psi_{0}|\hat{\psi}(\vec{x})|\Psi_{n}(N+1)\rangle,
\ \epsilon_{n}=E_{n}(N+1)-E_{0}\  \  \ for\  \  \epsilon_{n} \ge
\mu,
\end{equation}
or
\begin{equation}
\phi_{n}(\vec{x})=\langle\Psi_{n}(N-1)|\hat{\psi}(\vec{x})|\Psi_{0}\rangle,\
\epsilon_{n}=E_{0}-E_{n}(N-1)\  \  \ for\  \  \epsilon_{n} < \mu.
\end{equation}
The wavefunctions $|\Psi_{n}(N\pm 1)\rangle$ and energy levels
$E_{n}(N\pm 1)$ are for the $N\pm 1$ electronic systems. The
functions $\{\phi_{n}(\vec{x})\}$ are the ones of
$\it{single}$-particle coordinates, and are called the
$\it{particle}$ states for those defined by Eq. (6) $(\epsilon_{n}
\ge \mu)$, and the $\it{hole}$ states for those defined by Eq. (7)
$(\epsilon_{n} < \mu)$, where $\mu$ is the chemical potential. The
corresponding energy $\epsilon_{n}$ are the electron affinity or
the electron ionization potential, respectively. A very important
feature of these single-particle states $\{\phi_{n}(\vec{x})\}$ is
that they form a $\it{complete}$ set as shown below,
\begin{equation}
\sum_{n}\phi_{n}(\vec{x})\phi^{*}_{n}(\vec{x}^{'})=\delta(\vec{x}-\vec{x}^{'}),
\end{equation}
where $n$ is for $\it{all}$ the hole or particle states. The Eq.
(5) is called the $\it{Lehmann}$ representation.

Define the average classical $\it{Coulomb}$ potential by
\begin{equation}
  V(\vec{x}) = \int
  v(\vec{x},\vec{x}^{'})\rho(\vec{x}^{'})d\vec{x}^{'},
\end{equation}
where
\begin{equation}
  \rho(\vec{x})
  =\langle\Psi_{0}|\hat{\psi}^{+}(\vec{x})\hat{\psi}(\vec{x})|\Psi_{0}\rangle,
\end{equation}
is the one-electron probability density, then the $\it{Green}$
function in the energy domain satisfies the following equation,
\begin{equation}
\left\{\epsilon-h(\vec{x})-V(\vec{x})\right\}G(\vec{x},\vec{x}^{'};\epsilon)-
\int
\Sigma(\vec{x},\vec{x}^{"};\epsilon)G(\vec{x}^{"},\vec{x}^{'};\epsilon)d\vec{x}^{"}
=\delta(\vec{x}-\vec{x}^{'}),
\end{equation}
where the operator $\Sigma(\vec{x},\vec{x}^{'};\epsilon)$ is
called the self-energy operator which is nonlocal and energy
dependent. From this equation for the single-particle $\it{Green}$
function and its $\it{Lehmann}$ representation (5), we can get an
equation that the single-particle states $\{\phi_{n}(\vec{x})\}$
satisfy
\begin{equation}
\left\{h(\vec{x})+V(\vec{x})\right\}\phi_{n}(\vec{x})+\int\Sigma(\vec{x},\vec{x}^{'};
\epsilon_{n})\phi_{n}(\vec{x}^{'})d\vec{x}^{'}=\epsilon_{n}\phi_{n}(\vec{x}),
\end{equation}
or
 \begin{equation}
\left\{h + V + \Sigma(\epsilon_{n})\right\}|\phi_{n}\rangle =
\epsilon_{n}|\phi_{n}\rangle
\end{equation}
in a more general $\it{Dirac}$ notation. It is called the
$\it{Dyson}$ equation or the energy eigenequation for the
quasi-particles in the current
literature~\cite{hedin,louie,sham,martin}. When we do the
comparison with the $\it{Hartree-Fock}$ equation or the
$\it{Kohn-Sham}$ equation~\cite{hartree,fock,kohn}, it seems that
the self-energy operator $\Sigma$ is related to the exchange and
correlation effects of an interacting many-electron system beyond
that of the classical $\it{Coulomb}$ interaction. Unlike the
$\it{Kohn-Sham}$ equation, however, where the explicit analytical
potential for the exchange-correlation potential is unknown, the
self-energy operator has intrinsic structure, and, for example,
can be expanded as a perturbation series as follows,
\begin{equation}
\Sigma = \Sigma^{(0)}+\Sigma^{(1)}+...+\Sigma^{(n)} +....
\end{equation}
They have explicit physical interpretations and therefore can be
approached in a systematic
way~\cite{hedin,louie,sham,martin,goscinski,ohrn1,simons,ohrn2,cederbaum,freed}.
Another important feature of Eq. (12) is that the single-particle
states are defined for both hole state (Eq. (6)) and particle
states (Eq. (7)), and therefore there exists the concept of a
fundamental excitation in the present formalism. In other words,
we can form the configurations based on these single particle
states. Furthermore, since they constitute a complete set of
single-particle states, as shown in Eq.(8), any $N$-electron
wavefunctions can be expanded as a linear combination of these
configurations. For these reasons, we can regard the equation (12)
as a most general eigenequation for creating the single-particle
states or the atomic or molecular orbitals at present time. It is
the corresponding $\it{one}$-particle description of an $N$
interacting $\it{many}$-body system~\cite{wei2}. For clearness and
easiness to be understood, we call the single-particle states
determined by Eq. (12) as the $\it{correlated}$ atomic or
molecular orbitals in order to make a distinction from those
determined from the $\it{Hartree-Fock}$ equation. Obviously, they
will catch the full Hamiltonian (2) more than do the
$\it{Hartree-Fock}$ orbitals.

The successfulness for obtaining the most appropriate correlated
atomic or molecular orbitals $\{\phi_{n}(\vec{x})\}$ will depend
on how well we can obtain the correct self-energy operator
$\Sigma$. This will in turn depend on what kind of wavefunctions
or what level of theories we select as the reference or the
initial wavefunction for our construction of $\Sigma$ since the
Eq. (12) is an $\it{iterative}$ equation for determination of
$\{\epsilon_{n}\}$ and $\{\phi_{n}(\vec{x})\}$. Obviously, there
will be different choices for different species or for different
molecular geometries being studied as have already been
demonstrated in many existing quantum chemistry calculations.
Several types of perturbation schemes for the self-energy operator
have already been developed either from solid state physics
community or by quantum
chemists~\cite{hedin,louie,sham,martin,goscinski,ohrn1,simons,ohrn2,cederbaum,freed}.
These include the functional derivative
method~\cite{hedin,louie,sham,martin}, the superoperator formalism
~\cite{goscinski,ohrn1,simons,ohrn2}, the diagrammatic expansion
method~\cite{cederbaum}, and the equation of motion
approach~\cite{freed}.

\vspace{0.35in}
\section{Calculation and Results}

In this section, we present the computation of the single-particle
properties and total energies for $H_{2}O$ molecule. We employ the
$\it{Hartree-Fock}$ method, $\it{DFT}$, and correlated molecular
orbital approach we describe above for the calculation and do the
corresponding comparison.

The geometric parameters for the water molecule are taken from
experimental observation which are $R(O-H)=0.957 \dot{A}$, and
$\angle HOH = 104.5 (deg)$~\cite{water1}. For the
$\it{Hartree-Fock}$ calculation, we use the $\it{cc-pVTZ}$ basis
set~\cite{basis1}. The calculated energies for the first ten
molecular orbitals are listed in the second column of Table 1. The
computed total energy is shown in the Table 2.
 For the $\it{DFT}$ calculation, we use the same set of basis
 functions. The exchange-correlation functional is approximated
 with the $\it{B3LYP}$ scheme~\cite{becke1,parr2}.
 The result for the first ten $\it{Kohn-Sham}$ orbital energies is listed in the
 third column of the Table 1. The total energy is shown in the
 Table 2.
For the computation based on the correlated molecular orbitals, we
take the second-order approximation for the self-energy operator,
\begin{equation}
 \Sigma_{ij}(E) = \Sigma^{(1)}_{ij}(E) + \Sigma^{(2)}_{ij}(E).
\end{equation}
 The detailed forms for the self-energy operator with different
 orders are dependent upon the reference states chosen~\cite{louie,wei3}.
For the closed-shell molecules, if we pick the $\it{Hartree-Fock}$
orbitals as the reference states, the first-order self energy
vanishes, and the second-order self-energy is given
by~\cite{szabo}
 \begin{equation}
  \Sigma^{(2)}_{ij}(E)= \sum_{ars}^{N/2} \frac{\langle
  rs|ia\rangle\left(2\langle ja|rs\rangle - \langle aj|rs\rangle
  \right)}{E+\epsilon_{a}-\epsilon_{r}-\epsilon_{s}} +
 \sum_{abr}^{N/2} \frac{\langle
  ab|ir\rangle\left(2\langle jr|ab\rangle - \langle rj|ab\rangle
  \right)}{E+\epsilon_{r}-\epsilon_{a}-\epsilon_{b}}
\end{equation}
where $a,b,...$ are the spatial hole states, and $r,s,...$ are the
spatial particle states. If we choose the $\it{Kohn-Sham}$
orbitals as the reference states, however, the first-order
self-energy takes the form
 \begin{equation}
  \Sigma_{ij}^{(1)}(E) = - \langle i|V_{xc}|j\rangle -
  \sum_{a}^{N} \langle ia|aj\rangle ,
\end{equation}
 and the second-order self-energy remains the same as that for the
 case of the $\it{Hartree-Fock}$ orbitals.
 We solve the eigenequation (12) for the quasi-particles
with the $\it{cc-pVTZ}$ basis set. When the $\it{Hartree-Fock}$
orbitals are used as the reference state, the calculated
quasienergies for the first ten correlated molecular orbitals are
shown in the third column of the Table 1. The resulting total
energy with the single determinant using the first five
doubly-occupied hole states is also listed in the table 2. When
the $\it{DFT}$ determinant is employed as the reference state, the
corresponding results are listed in the forth column of table 1 or
table 2. All the computations are done with the $\it{Hondo-v99.6}$
suite~\cite{truhlar}.

\vspace{0.35in}
\section{Discussion and Conclusions}

 In this paper, we present a novel approach for the study of electronic structure of atoms and molecules
  related to the single-particle Green function theory. We argue that the single-particle Green function
 provides a most general theoretical framework for generating the atomic or molecular
 orbitals for the atoms and molecules. Based on this statement, we have calculated both the energies of these
 single-particle states and total energies for the $H_{2}O$ molecule~\cite{karplus,jordan,goddard}. For the total energy, a single-determinant
 wavefunction composed of hole states only is used for the computation. At the same time, the calculations
 are also performed with the $\it{Hartree-Fock}$ and $\it{DFT}$ methods.

 When compared with the experimental ionization energy or electron
 affinity for $H_{2}O$ molecule~\cite{iwata}, we see that the correlated
 molecular orbitals with the $\it{Hartree-Fock}$ orbitals as the
 reference state gives the better results than the ones from the
 $\it{Hartree-Fock}$ or $\it{DFT}$ methods. The total energies obtained with three different methods are also
 compared to the one obtained from the experimental observation~\cite{water1,karplus,jordan,goddard}. The
 correlated molecular orbital approach results in the best value. Of course, the calculation can be further
 improved by choosing the $\it{DFT}$ as a reference wavefunction. We have the similar conclusion.

 Since the work of $\it{Heitler}$ and $\it{London}$ in the calculation of the
electronic structure for $H_{2}$ molecule, which is the indication
of the beginning of the field of quantum chemistry, it has the
history of development for more than eighty years. However, there
is a fundamental issue, i.e., the quality of atomic or molecular
orbitals, which has been neglected for a long time. This paper
addresses this "quality" issue for single-particle states or
orbitals in many-body theory. From the perturbation theory point
of view, this corresponds to a definition of the reference
Hamiltonian, which is crucial in the minimization of dynamic
correlation energy or convergence of perturbation series. It is
also critical in providing the best single particle properties.
Both of the calculated single-particle properties and total
energies have explicit physical interpretation and are subject to
the test from experimental observations~\cite{koopmans}.

 From above analysis, it is obvious that when the concept of correlated atom or molecular orbital is
 incorporated into the quantum many-body perturbation or coupled cluster theory, it will provide a most
 powerful quantum many-body approach for the study of electronic structure of atoms or
 molecules. On one hand, its single-particle properties have
 obvious physical meanings which is in contrast to the case for
 the $\it{DFT}$. Furthermore, it can go beyond the single-determinant
 level and form configurations. Therefore, it can study the issues
 when configuration mixing is important. On the other hand, when
 doing the comparison to the traditional quantum many-body theory
 based on the $\it{Hartree-Fock}$ or $\it{MCSCF}$ orbitals, the correlated
 orbital method not only has provided a better description of
 single-particle properties, but also gives us the better
 convergence at the configuration level and therefore provides a more
 powerful computational scheme. For these reasons, we could claim
 that the correlated atomic or molecular approach will be a most general
 $\it{ab}$ $\it{initio}$ correlation method for electron structure
 calculations. It possesses the advantages and also overcomes their shortcomings of current
 $\it{DFT}$ and conventional correlation approaches based on the atomic or molecular
orbitals determined from the $\it{Hartree-Fock}$ or $\it{MCSCF}$.

  Of course, it has been a very difficult task for a long time to get the
  approximate self-energy operator to the higher orders. However,
  the $\it{intrinsic}$ structure such as its perturbation series
  expansion has offered us a possibility instead of an $\it{outside}$ model for the approximation.
  Furthermore, the further study of this underlying intrinsic structure will tell us more universal
  things which might be true even for a many-body theory or system in
  general. Henceforth, the continuing investigation of the higher
  order self-energy operators and their relations will be a
  rewarding research~\cite{uccirati}.

  An interesting point needed to be mentioned is that the self-energy operator in Eq. (12) does not
  have to be Hermitian which corresponds to the situation when $\psi_{n}(\vec{x})$ is a real orbital.
  Here the imaginary case for the operator is related to the electron dynamics which is left as
  a future investigation~\cite{hedin}

  Finally, if we fully explore the usefulness of the pseudopotential theory, combined $\it{QM/MM}$ approach, or
  linear scaling algorithms and so forth, the correlation theory based on the correlated atomic or molecular
  orbitals will provide to us a most robust approach for the study of electronic structure even for large
  systems~\cite{wei1}.

\newpage

  \begin{center}
{\bf Table Caption}
\end{center}

\mbox{}\\
\noindent {\bf Table 1.} The single-particle properties or orbital
energies (in a.u.) of $H_{2}O$ molecule from the calculations
based on the $\it{Hartree-Fock}$, $\it{DFT}$ and $\it{correlated}$
molecular orbital approaches as well as from the experimental
measurement.

%\mbox{}\\
%\noindent {\bf Table 2.} The single-particle properties of $N_{2}$
%molecule from the calculations based on the $\it{Hartree-Fock}$,
%$\it{DFT}$ and $\it{correlated}$ molecular orbital approaches as
%well as from the experimental measurement.

 \mbox{}\\
\noindent {\bf Table 2.} The total energies (in a.u.) of $H_{2}O$
molecule from the calculations based on the $\it{Hartree-Fock}$,
$\it{DFT}$ and $\it{correlated}$ molecular orbital approaches as
well as from the experimental measurement or $\it{CI}$
calculation.

\end{document}